\documentclass[aps,prb,twocolumn,superscriptaddress,showpacs]{revtex4}

\usepackage{graphicx}
\usepackage{amsmath}
\usepackage{bm}

\renewcommand{\vec}[1]{\bm{#1}}

\begin{document}

\title{Effective field theory for $S=1$ quantum nematic}

\author{B. A. Ivanov} 
\email{bivanov@i.com.ua}
\affiliation{Institute of Magnetism, National Academy of Sciences and
Ministry of Education, 36(b) Vernadskii avenue, 03142 Kiev, Ukraine }

\author{A. K. Kolezhuk}
\homepage{http://www.itp.uni-hannover.de/~kolezhuk}
\affiliation{Institute of Magnetism, National Academy of Sciences and
Ministry of Education, 36(b) Vernadskii avenue, 03142 Kiev, Ukraine }
\affiliation{Institut f\"ur Theoretische Physik, Universit\"at
Hannover, Appelstra{\ss}e 2, 30167 Hannover, Germany}

\date{\today}

\begin{abstract}
For $S=1$ system with general isotropic nearest-neighbor exchange, we
derive the low-energy description of the spin nematic
phase in terms of the $\rm RP^{2}$ 
nonlinear $\sigma$-model.  In one dimension, quantum fluctuations
destroy long-range nematic (quadrupolar) ordering, leading to the
formation of a gapped spin liquid state being an analog of the Haldane
phase for a spin nematic.  Nematic analog of the Belavin-Polyakov instanton with
$\pi_{2}$ topological charge $1/2$ is
constructed. In two dimensions the long-range order is
destroyed by thermal fluctuations and at finite temperature the system
is in a renormalized classical regime. Behavior in external magnetic field is
discussed.
\end{abstract}
\pacs{75.10.Jm, 75.40.Cx, 75.40.Gb}

\maketitle

Low-dimensional spin systems have been attracting permanent attention
of researchers over more than half a century. A rich palette of their
physical properties determined by the essential role played by
quantum fluctuations makes them a very attractive
playground for testing various theoretical concepts. In the last two
decades, this interest has got a considerable impact, motivated
particularly by the increasing availability of quasi-low-dimensional
magnetic materials. A number of exotic ``quantum spin liquid'' states
has been discovered, the most wide known example being the famous
Haldane phase in integer-spin antiferromagnetic (AF) chains.
\cite{Haldane83}

A generic example of the Haldane phase is the isotropic Heisenberg 
spin-1 AF chain. However, the most general isotropic exchange interaction
for spin $S=1$ includes biquadratic terms as well,
which naturally leads to the model described by the following
Hamiltonian:
\begin{equation} 
\label{ham} 
\widehat{H}=\sum_{<\vec{n}\vec{\delta}>} 
\cos\theta\,
(\vec{S}_{\vec{n}\vphantom{\vec{\delta}}}\cdot\vec{S}_{\vec{n}+\vec{\delta}})  
+\sin\theta\,
(\vec{S}_{\vec{n}\vphantom{\vec{\delta}}}\cdot\vec{S}_{\vec{n}+\vec{\delta}})^{2},
\end{equation} 
where $\vec{S}_{\vec{n}}$ are spin-1 operators at the lattice site $\vec{n}$, and
summation over the nearest neighbors is implied. There are
indications\cite{Millet+99,Lou+00}
 that  moderate 
biquadratic exchange is
present in the quasi-one-dimensional compound $\rm
LiVGe_{2}O_{6}$. The points $\theta=\pi$ and $\theta=0$ correspond to the Heisenberg
ferro- and antiferromagnet, respectively. In one dimension (1D), the model
(\ref{ham}) is studied rather extensively, and a number of analytical and numerical
results for several particular cases are available.
\cite{Haldane83,Affleck86,Oitmaa+86,BloteCapel86,Solyom87,SinghGelfand88,Chang+89,%
Papanicolaou88,Chubukov90-91,%
ItoiKato97,FathSolyom93a,%
Uimin70Lai74Sutherland75,Takhtajan82Babujian82-83Kulish+81,%
Parkinson88,Klumper89-90,BarberBatchelor89,AKLT} It is firmly established that the
Haldane phase with a finite spectral gap occupies the interval $-\pi/4 <\theta <
\pi/4$, and the ferromagnetic state is stable for $\pi/2 < \theta< 5\pi/4$, while
$\theta=5\pi/4$ is an SU(3) symmetric point with highly degenerate ground
state.\cite{Batista+02}

Exact solution is available \cite{Uimin70Lai74Sutherland75} for the
Uimin-Lai-Suther\-land (ULS) point $\theta=\pi/4$ which has SU(3)
symmetry. The ULS point was
shown \cite{ItoiKato97} to mark the
Berezinskii-Kosterlitz-Thouless (BKT) transition from the massive
Haldane phase into a massless phase occupying the interval
$\pi/4<\theta<\pi/2$ between the Haldane and ferromagnetic phase; this
is supported by numerical studies.\cite{FathSolyom93a}

The properties of the remaining region between the Haldane and
ferromagnetic phase are more controversial.  The other Haldane phase
boundary $\theta=-\pi/4$ corresponds to the exactly solvable
Takhtajan-Babujian model; \cite{Takhtajan82Babujian82-83Kulish+81} the
transition at $\theta=-\pi/4$ is of the Ising type and the ground
state at $\theta<-\pi/4$ is spontaneously dimerized with a finite gap
to the lowest excitations.
\cite{Affleck86,BloteCapel86,Oitmaa+86,Solyom87,SinghGelfand88,Chang+89,FathSolyom93a}
The dimerized phase extends at least up to and over the point
$\theta=-\pi/2$ which has a twofold degenerate ground state and finite
gap.  \cite{Parkinson88,Klumper89-90,BarberBatchelor89}

 Chubukov \cite{Chubukov90-91} used the Holstein-Primakoff-type
bosonic representation of spin-1 operators\cite{Papanicolaou88} based
on the quadru\-polar ordered ``spin nematic'' reference state with
$\langle \vec{S}\rangle=0$, $\langle S_{x,y}^{2}\rangle=1$, $\langle
S_{z}^{2}=0\rangle$, and suggested, on the basis of the
renormalization group (RG) arguments, that the region with
$\theta\gtrsim5\pi/4$ is a disordered nematic phase.  Early numerical
studies\cite{FathSolyom95} seemed to have ruled out this possibility,
and a common belief now \cite{SchadschneiderZittartz95,Katsumata95} is
that the dimerized phase extends all the way up to the ferromagnetic
phase, i.e., that it exists in the entire interval
$5\pi/4<\theta<7\pi/4$. However, recent numerical results
\cite{Kawashima02} indicate that the dimerized phase ends at certain
$\theta_{c}>5\pi/4$, casting doubt on the conclusion reached nearly a
decade ago.

The aim of the present paper is to show that the low-energy dynamics
of the model (\ref{ham}) for $\theta\gtrsim5\pi/4$ can be effectively
described by the nonlinear $\sigma$ model for a unit \emph{director}
field (i.e., a unit vector whose opposite directions are physically
identical).  The coupling constant becomes small in the vicinity of
the ferromagnetic phase boundary $\theta=5\pi/4$.  This formulation
allows one to establish many properties of the nematic phase by using
extensive results available for the standard (vector field) $O(3)$
nonlinear $\sigma$ model.  We also study the effect of external
magnetic field, which can be easily incorporated in our formalism.  We
argue that in 1D the ground state is disordered, in a complete analogy
with the Haldane phase in case of the vector $O(3)$ model, and its
elementary excitation is a massive triplet. In 2D, long-range nematic
order exists only at $T=0$. An explicit solution for the
Belavin-Polyakov instanton with \emph{half-integer} charge in a
(1+1)-dimensional isotropic nematic is presented.

We start by introducing the following set of coherent states for
$S=1$:
\begin{equation} 
\label{cs}
|\vec{u},\vec{v}\rangle =\sum_{j}(u_{j}+iv_{j})|t_{j}\rangle, \quad
 j\in (x,y,z), 
\end{equation}
where $|t_{j}\rangle$ are three  ``cartesian'' spin-1 states:
\begin{equation} 
\label{t}
|\pm\rangle =\mp(1/\sqrt2)(|t_{x}\rangle\pm i|t_{y}\rangle),\quad
|0\rangle=|t_{z}\rangle. 
\end{equation} 
The coherent state is characterized by vectors $\vec{u}$ and $\vec{v}$ satisfying the
normalization constraint $\vec{u}^{2}+\vec{v}^{2}=1$. The freedom to choose an
overall phase factor can be fixed by setting $\vec{u}\cdot \vec{v}=0$. It is easy to
check that the resolution of identity $\frac{3}{4\pi^{2}}\int {\cal
D}(\vec{u},\vec{v}) |\vec{u},\vec{v}\rangle \langle \vec{u},\vec{v}|=1 $ holds.

In what follows we are interested in the region around $\theta=5\pi/4$, hence it is
convenient to use the notation 
\[
\cos\theta\equiv -J_{1},\quad \sin\theta\equiv -J_{2},\qquad J_{2}\gtrsim J_{1}>0
\] 
Using the states (\ref{cs}), one can construct the
coherent state path integral, and the effective Lagrangian will have
the form
\begin{subequations}
\label{Leff}
\begin{equation} 
\label{Ldyn}
L_{\rm eff}=-2\hbar \sum_{\vec{n}} \vec{v}_{\vec{n}}\cdot
\partial_{t}\vec{u}_{\vec{n}}
-\sum_{<\vec{n}\vec{\delta}>} \langle \widehat{h}_{\vec{n},\vec{n}+\vec{\delta}}\rangle,
\end{equation}
where the average of the local Hamiltonian for two neighboring sites $1$ and $2$
is, up to a constant, given by
\begin{eqnarray}  
\label{Hav}
\langle \widehat{h}_{12}\rangle&=&-4J_{1}\big\{ (\vec{u}_{1}\cdot\vec{u}_{2})
 (\vec{v}_{1}\cdot\vec{v}_{2}) - (\vec{u}_{1}\cdot\vec{v}_{2})
 (\vec{v}_{1}\cdot\vec{u}_{2})\big\}\nonumber\\
&-&J_{2}\big\{(\vec{u}_{1}\cdot\vec{u}_{2}-\vec{v}_{1}\cdot \vec{v}_{2})^{2}
+(\vec{u}_{1}\cdot\vec{v}_{2}+\vec{v}_{1}\cdot
 \vec{u}_{2})^{2}\big\}\nonumber\\
&-&\vec{B}\cdot (\vec{u}_{1}\times\vec{v}_{1}+\vec{u}_{2}\times\vec{v}_{2}). 
\end{eqnarray}
\end{subequations} 
Here we have included the Zeeman term
$-\vec{B}\cdot\sum_{\vec{n}}\vec{S}_{\vec{n}}$ describing external
magnetic field $\vec{B}$.

\begin{figure}[tb]
\includegraphics[width=70mm]{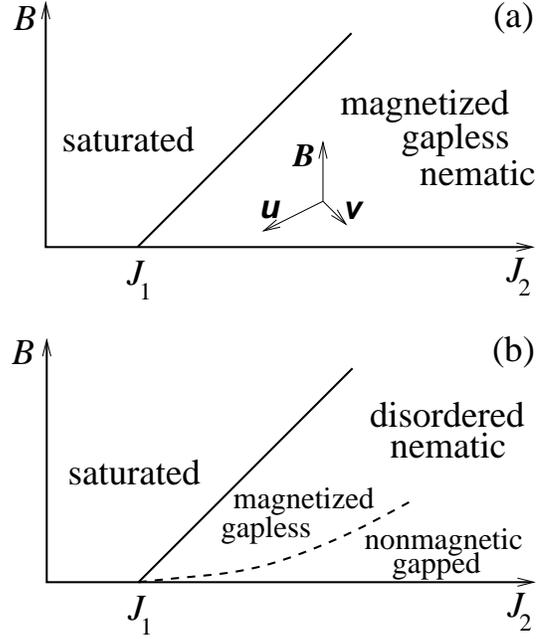}
\caption{\label{fig:ndiag} Schematic $T=0$ phase diagram
of the model (\protect\ref{ham}) in the vicinity of the critical point
$J_{2}=J_{1}$: (a) in dimension $D\geq 2$; (b) in one dimension. }
\end{figure}

  Assuming a uniform ground state and minimizing $\langle \widehat{H}\rangle$, one
arrives at the mean-field phase diagram shown in Fig.\
\ref{fig:ndiag}(a): at zero field the ferromagnetic phase with $u=v=1/\sqrt{2}$ is
stable for $J_{2}<J_{1}$, and at $J_{2}=J_{1}$ one has a degenerate first-order
transition into the nematic phase with $\vec{v}=0$ and parallel alignment of
$\vec{u}$ (actually, $\vec{u}$ and $\vec{v}$ can be used on equal terms, and we just
voluntarily choose $\vec{v}$ to be zero in the ground state). Note that vector
$\vec{u}$ is in this case a \emph{director} since $\vec{u}$ and $-\vec{u}$ correspond
to the same state. In external field the system acquires finite magnetization
$\langle \vec{S}\rangle=\vec{B}/[Z(J_{2}-J_{1})]$, where $Z$ is the coordination
number of the lattice, while nematic order persists in plane perpendicular to
$\vec{B}$.  The magnetization increases with the field, and at $B=Z(J_{2}-J_{1})$ the
nematic undergoes a second order phase transition into the phase with fully saturated
magnetic moment.

Our next aim is to study how the above classical mean-field picture changes due to 
quantum or thermal fluctuations.
We pass to the continuum limit in (\ref{Leff}), viewing $\vec{u}$
and $\vec{v}$ as smooth field distributions.
From the mean-field solution one may assume that $v\ll u$,
and from the form of the Lagrangian (\ref{Ldyn}) it is clear that
$\vec{v}$ plays the role of momentum conjugate to $\vec{u}$, so that
$\vec{v}$ will be eventually proportional to the time derivative of $\vec{u}$
(later this will be checked in a self-consistent
way). We will thus keep only terms up to the second order in $\vec{v}$,
and derivatives of $\vec{v}$ will be neglected. Doing so, one obtains
the following continuum version of the Lagrangian:
\begin{eqnarray} 
\label{Luv} 
L[\vec{u},\vec{v}]&=& V_{0}^{-1}\int d^{D}r \Big\{ 
-2\hbar\,\vec{v}\cdot \partial_{t}\vec{u} -2Z(J_{2}-J_{1})\vec{u}^{2}
\vec{v}^{2}\nonumber\\
&+&2\vec{v}\cdot(\vec{B}\times \vec{u}) -(J_{2}/2)\sum_{\alpha=1}^{Z} \big(
(\vec{\delta}_{\alpha}\cdot \vec{\nabla})\,\vec{u}\big)^{2}
\Big\},
\end{eqnarray}
where $V_{0}$ is the volume of the elementary cell of the
$D$-dimensional lattice,
$\vec{\delta}_{\alpha}$ are vectors describing the position of $Z$
nearest neighbors with respect to a given lattice site, and
constraints $\vec{u}^{2}+\vec{v}^{2}=1$, $\vec{u}\cdot \vec{v}=0$ are
implied. In what follows, we will assume for simplicity that the
lattice is hypercubic, then $Z=2D$, $V_{0}=a^{D}$, and
$(\vec{\delta}_{k}\cdot\vec{\nabla})=a\nabla_{k}$,
$k=1\ldots D$, where $a$ is the lattice constant. 

The ``slave'' variable $\vec{v}$  under the assumption $v\ll u$ can be integrated out,
yielding
\begin{equation} 
\label{v} 
\vec{v}=[2Z(J_{2}-J_{1})]^{-1}\{(\vec{B}\times\vec{u})-\hbar\partial_{t}\vec{u}
\}.
\end{equation}
Substituting (\ref{v}) back into (\ref{Luv}) gives the following
effective Lagrangian depending on $\vec{u}$ only: 
\begin{equation} 
\label{Lu} 
L_{\rm eff}=\frac{J_{2}}{c^{2}}\int \frac{d^{D}r}{a^{D-2}}
\Big\{
\Big(\partial_{t}\vec{u}-\frac{\vec{B}\times\vec{u}}{\hbar}\Big)^{2}
-c^{2}\sum_{k=1}^{D}
(\nabla_{k}\vec{u})^{2} 
\Big\},
\end{equation}
where
$c=\big[2ZJ_{2}(J_{2}-J_{1})\big]^{1/2}a/\hbar$ is the characteristic limiting velocity,
and $\vec{u}$ now should be considered as a unit vector,
$\vec{u}^{2}=1$. 
Note that according to (\ref{v}) a change of sign of $\vec{u}$
automatically means a sign change for $\vec{v}$, so that $\vec{u}$ is in this
approximation completely equivalent to  $-\vec{u}$.
The above description  is valid at the energy scales
$E<E_{0}=2Z(J_{2}-J_{1})$.
One readily observes that (\ref{Lu}) is nothing but the Lagrangian of
the well-known
nonlinear $\sigma$ model used as the effective theory for antiferromagnets
\cite{Haldane83,Affleck89rev} (without the topological term). 
Even the additional terms in the second line of
Eq.\ (\ref{Lu}), describing the effect of the external magnetic field,
are identical to those appearing in the $\sigma$
model for antiferromagnets.
Thus, the low-energy dynamics of model (\ref{ham}) in the nematic
phase is similar to the dynamics of an antiferromagnet,
with the only yet \emph{important} difference that instead of the unit vector
of sublattice magnetization one now has the nematic director
$\vec{u}$: the order parameter space is $RP^{2}$ instead of
$S^{2}$. 
The $\sigma$-model formulation, in contrast to the spinwave approach of
Chubukov,\cite{Chubukov90-91} allows one to study full nonlinear dynamics of the
problem. 

At zero field, one can rewrite the  Lagrangian (\ref{Lu})
in a standard notation using dimensionless
space-time variables $x=(x_{0},\vec{x})$, $\vec{x}=\vec{r}/a$,
$x_{0}=ict/a$. The effective Euclidean action takes the
following compact form:
\begin{equation} 
\label{sigma} 
\frac{{\cal A}_{\rm E}}{\hbar}=\frac{1}{2g}\int 
\Big(\frac{\partial \vec{u}}{\partial x_{\mu}} \Big)^{2}\, d^{{D+1}}x\,,
\end{equation}
with the coupling constant $g$ is defined as
\begin{equation} 
\label{g} 
g=\left\{ Z(J_{2}-J_{1})/2J_{2}\right\}^{1/2}.
\end{equation}
Note that smallness of the coupling constant does not require a large-$S$
approximation, and is controlled solely by the closeness to the ferromagnetic phase
boundary.

\emph{In one dimension ($D=1$)} continuous symmetry cannot be broken, and the ground
state of (\ref{sigma}) is disordered with exponentially decaying correlations.  The
correlation length $\xi$ for the usual $O(3)$ (vector) $\sigma$ model
can be obtained within  Polyakov's RG approach
\cite{Polyakov75NelsonPelcovits77} as
$\xi_{O(3)}\sim a e^{2\pi/g}$. In the $RP^{2}$ $\sigma$-model,
however, there is a rescaling in flow equations  
because of the change in the measure:
 the physical field
is not $\vec{u}$, but the bilinear projector 
$R=\vec{u}^{T}\otimes\vec{u}$.
The action can be rewritten as 
\begin{equation} 
\label{sigma1} 
\frac{{\cal A}_{\rm E}}{\hbar}=\frac{1}{4g}\int 
\langle \partial_{\mu}R,\partial_{\mu} R\rangle\, d^{{D+1}}x\,, 
\end{equation}
where $\langle A,B\rangle=\mbox{tr}(A^{T}B)$ denotes the scalar
product. The $\beta$-function in the leading order is the same as for
the $O(3)$ model,\cite{Zinn-Justin-book} $\beta(\Gamma)=-\frac{1}{2\pi}\Gamma^{2}$,
with the trivially
rescaled coupling constant $\Gamma=2g$.
Thus, for the
correlation length in  the
$RP^{2}$ $\sigma$-model one obtains
\begin{equation} 
\label{xi-1D} 
\xi_{RP^{2}}\sim a e^{\pi/g}=ae^{\pi\sqrt{J_{2}/(J_{2}-J_{1})}},
\end{equation} 
in agreement with Chubukov's one-loop RG result \cite{Chubukov90-91}
for interacting spin waves.  The elementary excitation is a massive
spin-1 triplet, and the gap $\Delta=\hbar c/\xi$ opens up
exponentially slow as one moves away from the phase transition point
$J_{2}=J_{1}$:
\begin{equation} 
\label{gap} 
\Delta\sim 2[J_{2}(J_{2}-J_{1})]^{1/2} e^{-\pi\sqrt{J_{2}/(J_{2}-J_{1})}}.
\end{equation}

The $RP^{2}$ and $O(3)$ $\sigma$-models are also different with
respect to their topological excitations.  In the $O(3)$ model there is a
localized solution with nonzero $\pi_{2}$ topological charge
$Q=\frac{1}{8\pi}\int d^{2}\xi\, \varepsilon_{\mu\nu} \vec{u}\cdot
(\partial_{\mu} \vec{u}\times\partial_{\nu} \vec{u}) $, known as the
Belavin-Polyakov instanton (BPI). \cite{BelavinPolyakov75} The
simplest BPI with $Q=1$ is described by $w=(z-a)/(z-b)$, where the
complex variable $w(\vec{u})=(u_{1}+iu_{2})/(1-u_{3})$ is defined as a
function of the complex coordinate $z=x_{1}+ix_{0}$, and generally
\emph{any analytical function} $w(z)$ yields a
solution.\cite{BelavinPolyakov75} The BPI action
$\mathcal{A}_{BPI}=4\pi\hbar Q/g$ does not depend on the parameters
$a$, $b$ which can be interpreted as coordinates of elementary
entities, \emph{merons}, constituting a BPI.  It was
speculated,\cite{Affleck89rev,MorenoOrland99} that the correlation
length $\xi_{O(3)}\propto e^{\mathcal{A}_{BPI}/2\hbar}$ is related to
the concentration of merons.

\begin{figure}[tb]
\includegraphics[width=30mm,angle=-90]{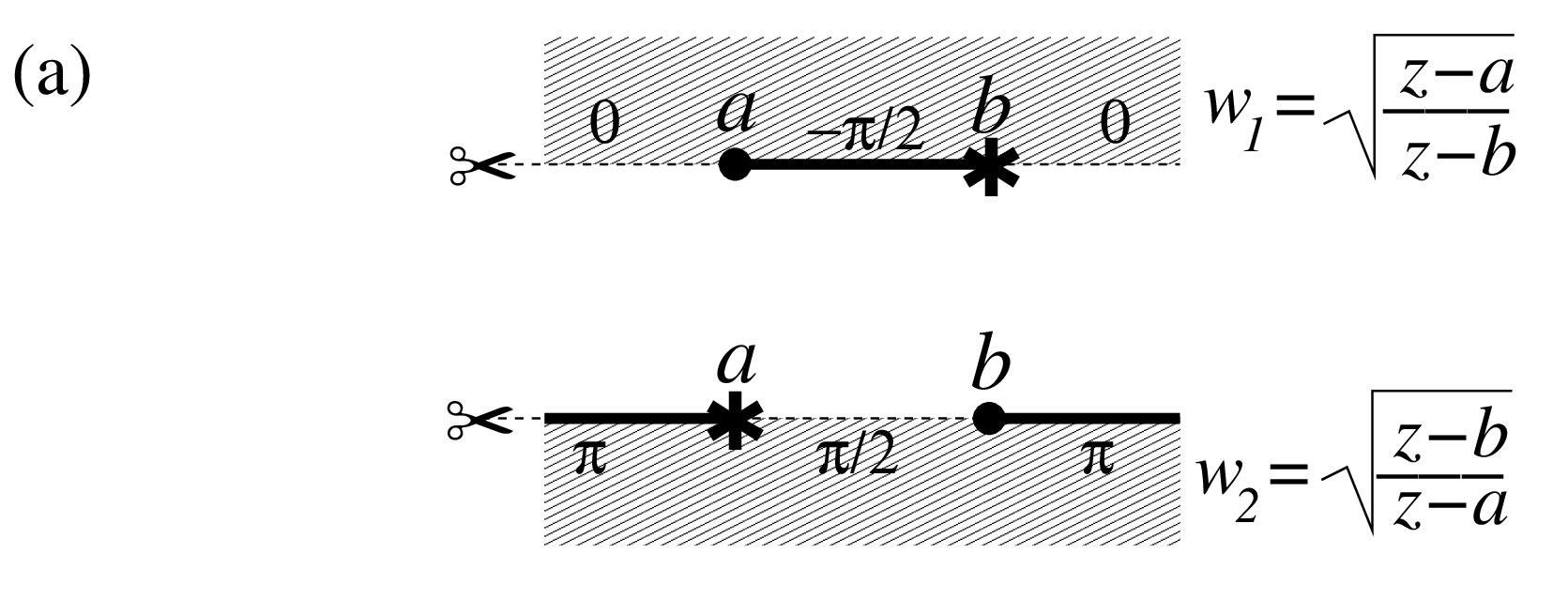}

\includegraphics[width=75mm]{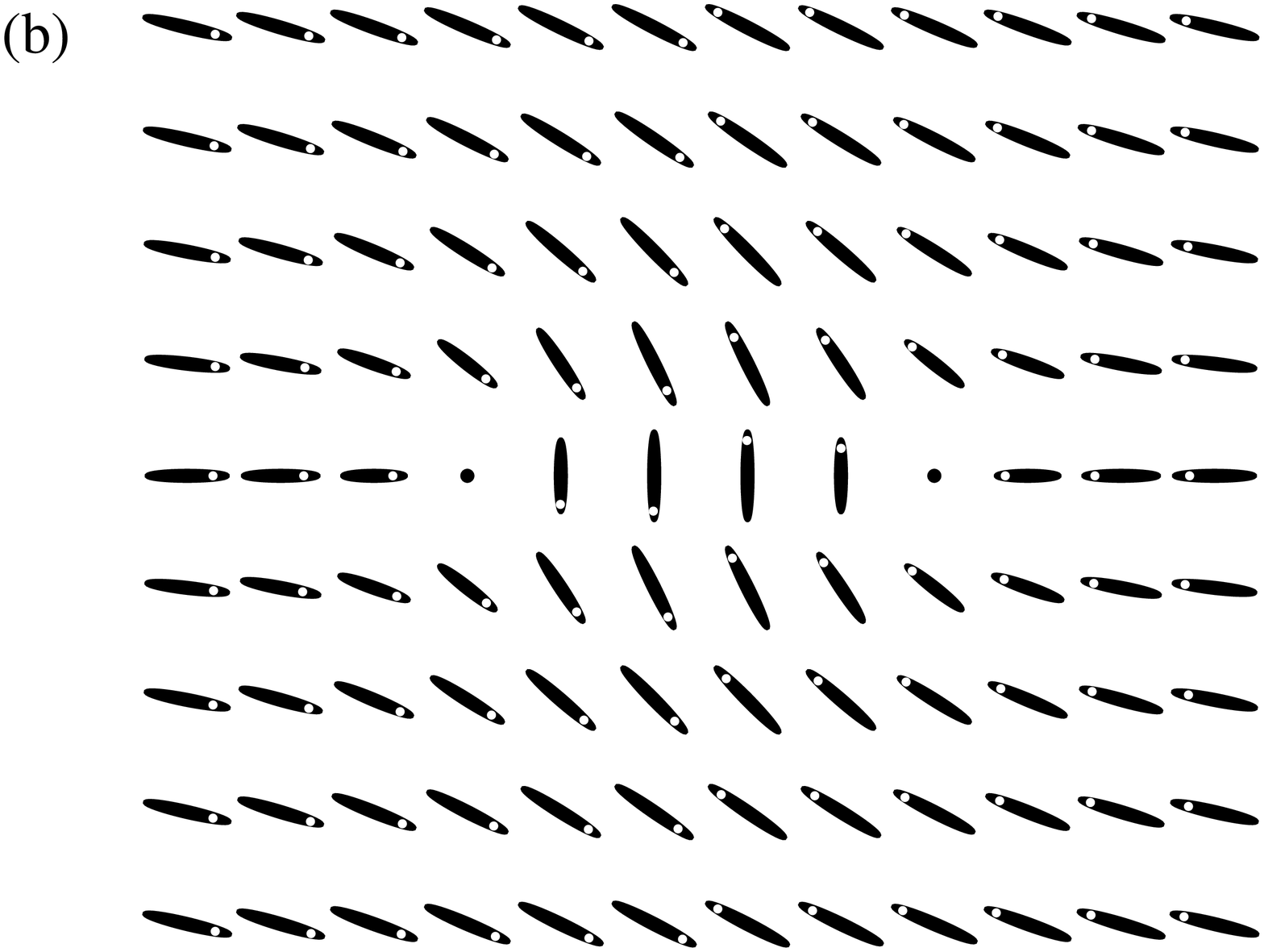}
\caption{\label{fig:bp-nem} Belavin-Polyakov-type soliton in a spin
nematic: 
(a) the solution is given by $w_{1}(z)$ in the
upper half-plane (respective to the line connecting points $a$ and
$b$), and by $w_{2}(z)$ in the lower half-plane, with the
appropriate branch cuts shown as thick solid lines and numbers denoting
the phase of  $w$; (b) schematic view of the solution, nematic
director $\vec{u}$ is  shown as an ellipsoid  in a
projection onto the figure plane; 
white spot indicates the end of ellipsoid
that is under the figure plane.
}
\end{figure}

In the $RP^{2}$ case the director
nature of the field makes possible BPI-type defects with
half-integer $Q$, whose action
is exactly one half of
that for their $O(3)$ $\sigma$ model counterparts:
Indeed, consider a solution of the form
$w=w_{1}(z)=\sqrt{\frac{z-a}{z-b}}$.
For the $O(3)$ $\sigma$-model  such
a solution would be invalid, because it has a branch cut. 
In a nematic, however,
$w(\vec{u})$ and $w(\vec{-u})$ are physically
identical, and the above solution can be matched with another one,
$w=w_{2}(z)=\sqrt{\frac{z-b}{z-a}}$,
so that
$w_{2}(\vec{u})=w_{1}(-\vec{u})$ 
on some line. This is easily achieved by choosing the cuts as
shown in Fig.\ \ref{fig:bp-nem}.
This solution has $Q=\pm\frac{1}{2}$,
and its action is just one half of that for the $O(3)$ BPI. Curiously,
this fact correlates with the extra factor
$\frac{1}{2}$ in the correlation length exponent (\ref{xi-1D}).

In the $RP^{2}$ model there is another type of topological defects, {\em disclinations},
characterized by a $\pi_{1}$ topological charge (vorticity) $q$. It is argued that
their presence could
produce the BKT transition \emph{in the isotropic
case}. \cite{KunzZumbach92} 
 However, one can see that such a transition
would occur above some critical value of the coupling $g_{BKT}$
of the order of $1$,
and as long as $g=(1-J_{1}/J_{2})^{1/2}\ll 1$, one may expect that the
disclinations will be bound in pairs and their effect can be neglected. 

Our approach easily allows one to incorporate the effect of an external magnetic
field.  Weak magnetic field $B<\Delta$ acts on the spectrum only in a trivial way
(the Zeeman shift), but at $B=\Delta$ the gap closes and the system enters the
critical phase with algebraically decaying correlations, which can be characterized
as the Tomonaga-Luttinger liquid.  \cite{Affleck91} The resulting phase diagram for
1D case is shown in Fig.\ \ref{fig:ndiag}(b).

\emph{For $D=2$,} at zero temperature the ground state should have
long-range nematic order, in agreement with recent numerical results,
\cite{HaradaKawashima02} as long as $g$ is small compared to some
finite value $g_{c}$ of the order of $1$ which marks the transition
into a quantum disordered phase. This latter transition is expected to
be the same as a finite-temperature transition in the
three-dimensional classical Lebwohl-Lasher model, which is the first
order supposedly due to the effect of disclination
lines.\cite{Priezjev01} At $T=0$ the phase diagram in presence of
magnetic field has the mean-field form of Fig.\ \ref{fig:ndiag}(a).
At $T\not=0$ and $B=0$ nematic order is destroyed by thermal
fluctuations, with the correlation length $\xi\sim a e^{2\pi J_{2}/T}$
(note extra factor $\frac{1}{2}$ in the exponent, as compared to the
standard result \cite{Polyakov75NelsonPelcovits77}).

In summary, we have shown that the low-energy dynamics of the bilinear-biquadratic
$S=1$ system (\ref{ham}) for $\theta\gtrsim5\pi/4$ can be effectively mapped onto the
$RP^{2}$ nonlinear $\sigma$ model.  We have argued that in one dimension this model
exhibits a disordered nematic state, supporting early proposition of Chubukov
\cite{Chubukov90-91} and recent numerical results \cite{Kawashima02} against the
commonly adopted\cite{FathSolyom95,SchadschneiderZittartz95,Katsumata95} point of
view.  Using parallels with the extensively studied vector version of the
$\sigma$-model, one can easily extract necessary information on the properties of
nematic phase. An instanton solution of the Belavin-Polyakov type with half-integer
topological charge is presented.

{\em Acknowledgments.--- }
This work is supported in part by the grant I/75895 from the
Volkswagen-Stiftung.


\begin{thebibliography}{10}

\bibitem{Haldane83} F.~D.~M.~Haldane, Phys. Lett. A {\bf 93}, 464
(1983); Phys. Rev. Lett. {\bf 50}, 1153 (1983).

\bibitem{Millet+99} P. Millet, F. Mila, F. C. Zhang, M. Mambrini,
A. B. Van Oosten, V. A. Pashchenko, A. Sulpice, and A. Stepanov,
Phys. Rev. Lett. {\bf 83}, 4176 (1999).

\bibitem{Lou+00} J. Lou, T. Xiang, and Z. Su, Phys. Rev. Lett {\bf 85}, 2380 (2000).

\bibitem{Affleck86} I. Affleck, Nucl. Phys. B {\bf 265}[FS15], 409
(1986); I. Affleck and F. D. M. Haldane, Phys. Rev. B {\bf 36}, 5291 (1987).

\bibitem{Oitmaa+86} J. Oitmaa, J. B. Parkinson, and J. C. Bonner,
J. Phys. C {\bf 19}, L595 (1986).

\bibitem{BloteCapel86} H. W. J. Bl\"ote and H. W. Capel, Physica A {\bf
139}, 387 (1986).

\bibitem{Solyom87} J. S\'olyom, Phys. Rev. B {\bf 36}, 8642 (1987).

\bibitem{SinghGelfand88} R. R. P. Singh and M. P. Gelfand,
Phys. Rev. Lett. {\bf 61}, 2133 (1988).

\bibitem{Chang+89} K. Chang, I. Affleck, G. W. Hayden, and Z. G. Soos,
J. Phys.: Condens. Matter {\bf 1}, 153 (1989). 

\bibitem{Papanicolaou88} N. Papanicolaou, Nucl. Phys. B {\bf 305} [FS23], 367 (1988).

\bibitem{Chubukov90-91} A. V. Chubukov, J. Phys.: Condens. Matter {\bf
2}, 1593 (1990); A. V. Chubukov, Phys. Rev. B {\bf 43}, 3337 (1991).



\bibitem{ItoiKato97} C. Itoi and M.-H. Kato, Phys. Rev. B {\bf 55},  8295 (1997).

\bibitem{FathSolyom93a} G. F{\'a}th and J. S{\'o}lyom, Phys. Rev. B {\bf 47}, 872
  (1993). 


\bibitem{Uimin70Lai74Sutherland75} G. V. Uimin, JETP Lett. {\bf 12},
225 (1970); C. K. Lai, J. Math. Phys. {\bf 15}, 1675 (1974);
B. Sutherland, Phys. Rev. B {\bf 12}, 3795 (1975).


\bibitem{Takhtajan82Babujian82-83Kulish+81} L. A. Takhtajan,
Phys. Lett. A {\bf 87}, 479 (1982); H. M. Babujian, Phys. Lett. A {\bf
90}, 479 (1982); Nucl. Phys. B {\bf 215}, 317 (1983); 
P. Kulish,
N. Reshetikhin, and E. Sklyanin, Lett. Math. Phys. {\bf 5}, 393
(1981).

\bibitem{Parkinson88} J. B. Parkinson, J. Phys. C {\bf 21}, 3793 (1988).

\bibitem{Klumper89-90} A. Kl\"umper, Europhys. Lett. {\bf 9}, 815
(1989); J. Phys. A {\bf 23}, 809 (1990); Int. J. Mod. Phys. B {\bf 4},
871 (1990).

\bibitem{BarberBatchelor89} M. N. Barber and M. T. Batchelor,
Phys. Rev. B {\bf 40}, 4621 (1989).

\bibitem{AKLT} I. Affleck, T. Kennedy, E. H. Lieb, and H. Tasaki,
Phys. Rev. Lett. {\bf 59}, 799 (1987); Commun. Math. Phys. {\bf 115},
477 (1988).

\bibitem{Batista+02} C. D. Batista, G. Ortiz, and J. E. Gubernatis, Phys. Rev. B {\bf
65}, 180402(R) (2002).


\bibitem{FathSolyom95} G. F{\'a}th and J. S{\'o}lyom, Phys. Rev. B
{\bf 51}, 3620 (1995).




\bibitem{SchadschneiderZittartz95} A. Schadschneider and J. Zittartz,
Ann. Physik {\bf 4}, 157 (1995).

\bibitem{Katsumata95} K. Katsumata, J. Mag. Magn. Mater. {\bf
140-144}, 1595 (1995) and references therein.

\bibitem{Kawashima02} N. Kawashima, Prog. Theor. Phys. Suppl. {\bf 145}, 138 (2002). 

\bibitem{Affleck89rev} I. Affleck, J. Phys.: Condens. Matter {\bf 1},
3047 (1989).


\bibitem{Polyakov75NelsonPelcovits77} 
A. M. Polyakov, Phys. Lett. B {\bf 59}, 79 (1975); 
D. R. Nelson and R. A. Pelcovits, Phys. Rev. B {\bf 16}, 2191 (1977).

\bibitem{Zinn-Justin-book} J. Zinn-Justin, \textit{Quantum Field
Theory and Critical Phenomena} (Oxford Univ. Press, 2002), \S15.6.

\bibitem{BelavinPolyakov75} A. A. Belavin and A. M. Polyakov, Pis'ma v
Zh. Eksp. Teor. Fiz. {\bf 22}, 503 (1975).

\bibitem{MorenoOrland99} E. Moreno and P. Orland, JHEP {\bf 04}, 002 (1999).

\bibitem{KunzZumbach92} H. Kunz and G. Zumbach, Phys. Rev. {\bf 46}, 662 (1992).

\bibitem{Affleck91}
I. Affleck, Phys. Rev. B {\bf 43}, 3215 (1991);
R. Konik and P. Fendley, preprint cond-mat/0106037.


\bibitem{HaradaKawashima02} K. Harada and N. Kawashima, Phys. Rev. B
{\bf 65}, 052403 (2002).



\bibitem{Priezjev01} N. V. Priezjev and R. A. Pelcovits, Phys. Rev. E {\bf 64},
  031710 (2001). 


\end{thebibliography}
\end{document}